\documentclass[twocolumn, deluxetables]{aastex631}
\usepackage{amsmath}
\usepackage{multirow}
\usepackage{graphicx}

\graphicspath{{./}{figures/}}

\shorttitle{XRISM Spectroscopy of M81*}
\shortauthors{Miller et al.}

\begin{document}

\title{XRISM Reveals a Remnant Torus in the Low-Luminosity AGN M81*}

\author[0000-0003-2869-7682]{Jon M. Miller}
\email{jonmm@umich.edu}
\affiliation{Department of Astronomy, University of Michigan, Ann Arbor, MI, 48109, USA}

\author[0000-0001-9735-4873]{Ehud Behar}
\affiliation{Department of Physics, Technion, Technion City, Haifa 3200003, Israel} 

\author{Hisamitsu Awaki}
\affiliation{Department of Physics, Ehime University, Ehime 790-8577, Japan} 

\author[0000-0001-8667-2681]{Ann Hornschemeier}
\affiliation{NASA / Goddard Space Flight Center, Greenbelt, MD 20771, USA}

\author{Jesse Bluem}
\affiliation{Johns Hopkins University, MD 21218, USA} 
\affiliation{NASA / Goddard Space Flight Center, Greenbelt, MD 20771, USA}

\author[0009-0006-4968-7108]{Luigi Gallo}
\affiliation{Department of Astronomy and Physics, Saint Mary's University, Nova Scotia B3H 3C3, Canada} 

\author[0000-0001-7773-9266]{Shogo B. Kobayashi}
\affiliation{Faculty of Physics, Tokyo University of Science, Tokyo 162-8601, Japan} 

\author[0000-0002-7962-5446]{Richard Mushotzky}
\affiliation{Department of Astronomy, University of Maryland, College Park, MD 20742, USA} 

\author{Masanori Ohno}
\affiliation{Department of Physics, Hiroshima University, Hiroshima 739-8526, Japan} 

\author[0000-0003-3850-2041]{Robert Petre}
\affiliation{NASA / Goddard Space Flight Center, Greenbelt, MD 20771, USA}

\author{Kosuke Sato}
\affiliation{Department of Physics, Saitama University, Saitama 338-8570, Japan} 
\affiliation{International Center for Quantum-field Measurement Systems for Studies of the Universe and Particles (QUP) / High Energy Accelerator Research Organization (KEK)}

\author{Yuichi Terashima}
\affiliation{Department of Physics, Ehime University, Ehime 790-8577, Japan} 

\author[0000-0001-6366-3459]{Mihoko Yukita}
\affiliation{Johns Hopkins University, MD 21218, USA} 
\affiliation{NASA / Goddard Space Flight Center, Greenbelt, MD 20771, USA}




\begin{abstract}
Up to 40\% of galaxies in the local universe host a low-luminosity active galactic nucleus (LLAGN), making it vital to understand this mode of black hole accretion.  However, the presence or absence of Seyfert-like geometries -- an accretion disk close to the black hole, an optical broad line region (BLR), and a molecular torus -- remains uncertain owing to the low flux levels of sources within this class.  Herein, we present an analysis of a XRISM/Resolve spectrum of M81*, the LLAGN in the heart of the nearby spiral galaxy M81.  A weak, neutral Fe~K emission line is detected and resolved into K$_{\alpha,1}$ and K$_{\alpha,2}$ components.  It shows a negligible velocity shift, and weak broadening (FWHM$=460^{+260}_{-160}~{\rm km}~{\rm s}^{-1}$) that corresponds to an inner emission radius of ${\rm r} \geq 2.7\times 10^{4}~GM/c^{2}$ for likely inclinations.  The Fe~K$_{\alpha}$ line likely traces a torus.
The upper limit on additional splitting of the Fe~K$_{\alpha}$ line components translates to a limit on the local magnetic field of ${\rm B} \leq 3.5\times 10^{8}$~Gauss, assuming Zeeman splitting. 
The spectra also reveal ionized plasma(s) through He-like Fe~XXV and H-like Fe~XXVI emission lines.  These can be fit equally well assuming photoionization and collisional excitation.  The H-like Fe~XXVI line is better described when a second component is included with a red-shift of ${\rm v} = 1600~{\rm km}~{\rm s}^{-1}$, but this addition is of marginal statistical significance.  We discuss these results in the context of radiatively inefficient accretion flow models, magnetically arrested disks, and possible links to the Fermi bubbles in the Milky Way.
\end{abstract}

\keywords{X-rays: black holes --- accretion -- accretion disks}


\section{Introduction}
While massive black holes accrete in a quasar or Seyfert mode, they can have profound impacts on their host galaxy or even on the structure of hot gas within clusters of galaxies (for a review, see \citealt{fabian2012}).  However, accretion at high fractions of the Eddington limit may be relatively short-lived compared to accretion at more modest rates (e.g., \citealt{hopkins2009}).  Surveys indicate that one-third of all galaxies and two-thirds of active galaxies can be classified as harboring low-ionization nuclear line emission regions (LINERs), powered by accretion at a small fraction of the Eddington limit \citep{ho2008}.   Given the prevalence of this accretion mode in the present universe, it is crucial to understand gas accretion and ejection in low-luminosity AGN (LLAGN).

Stellar-mass black holes in the ``low/hard'' state -- a low-luminosity phase observed in the decline phase of outbursts -- may provide a framework for understanding LLAGN (reviews, see, e.g., \citealt{rm2006}, \citealt{ho2008}).  In both cases, thermal emission from the inner disk is reduced by orders of magnitude compared to phases associated with high Eddington fractions, implying a similar reduction in the mass accretion rate.  Relativistic jets turn on in low-Eddington phases, again implying major changes in the central engine and accretion disk.  If AGN fueling follows the pattern observed in stellar-mass black holes -- typically, a rapid rise in $\dot{m}$ followed by an exponential decay over a much longer period -- then most LLAGN likely found in a slowly declining $\dot{m}$ phase rather than a rapid rise.  This should have two consequences for Seyfert geometries: (1) the BLR in LLAGN may be diminished and altered owing to a far weaker central engine radiating a different ionizing spectrum, and (2) the torus may be less visible in IR owing to reduced heating from central engine, but also due to a reduced covering factor as the mass reservoir has been depleted.  

M81 is a spiral galaxy that lies at a distance of ${\rm d} = 3.61^{+0.21}_{-0.19}$\,Mpc \citep{tully2016}; it has a small peculiar velocity of $z = -1.3\times 10^{-4}$ or $v = -39~{\rm km}~{\rm s}^{-1}$.  The mass of the black hole powering M81* (the nuclear black hole in M81) is constrained to be ${\rm M} = 7^{+2}_{-1}\times 10^{7}~{\rm M}_{\odot}$ via Hubble/STIS modeling of the broad H$\alpha$ line that is observed in this source \citep{devereux2003}.  Owing to its proximity, M81* is the brightest and best-studied example of its class, but the best observations still paint a confusing and sometimes self-contradictory picture.

Continuum flux properties of M81* appear to point 
to an accretion flow that must differ from that found in typical Seyferts.  Observations with the broad-band X-ray telescope (BBXRT; \citealt{serlemitsos1992}) measured a 0.6--10.0~keV X-ray flux of ${\rm F} = 3.6\times 10^{-11}~{\rm erg}~{\rm cm}^{-2}~{\rm s}^{-1}$ with a $\Gamma = 2.2$ power-law model \citep{petre1993}, corresponding to an X-ray luminosity of ${\rm L}_{\rm X} = 5.6\times 10^{40}~{\rm erg}~{\rm s}^{-1}$.  Combining Hubble and other telescopes, \citet{ho1996} noted that M81* lacks the ``big blue bump'' typical of disk accretion in Seyfert AGN.  Rather, the UV spectrum decreases with frequency in a $\nu {\rm L}_{\nu}$ sense, and the inferred bolometric luminosity is ${\rm L}_{\rm bol} = 9.3\times 10^{40}~{\rm erg}~{\rm s}^{-1}$ (giving ${\rm L}/{\rm L}_{\rm Edd} = 10^{-5}$).  Thus, the UV luminosity of M81* is at most comparable to the X-ray luminosity, in contrast to Seyferts wherein typical bolometric corrections range between 15--70 \citep{vasudevan2007}.  Whereas a torus is indicated in Seyferts by an IR rise, the MIR--NIR spectrum of M81* is relatively flat; this may be partly due to synchrotron flux from a jet \citep{mason2012}.  M81* is sometimes classified as radio--loud, and is observed to power a jet with discrete knot ejections typical of AGN in the radio mode \citep{king2016}. 

Two families of models -- radiatively inefficient accetion flows (RIAFs; \citealt{quataert1999}) and magnetically arrested disk models (MADs; \citealt{tchekhovskoy2011}) predict that a geometrically thin accretion disk does not extend to the innermost stable circular orbit at low Eddington fractions, and is instead truncated at larger radii.  MAD models also provide a natural explanation for jet production through magnetic fields.  The paradigm view that LLAGN are powered by RIAF-like flows is partly based on the broad-band properties of M81*.  However, the RIAF model for M81* was developed assuming a black hole mass of ${\rm M} = 4\times 10^{6}~{\rm M}_{\odot}$, and \citet{quataert1999} explicitly note the following: ``if the true mass were larger by a factor of $>10$, the thin disk emission would be colder (${\rm T} \propto m^{-1/4}$) and would therefore be capable of explaining the observations (we find, e.g., that $m = 4\times 10^{7}$, $\dot{m} = 10^{-5}$, and $r_{in} = 3$ can account for the observed optical/UV spectrum).''  (Here, the mass has units of solar masses, ${\rm M}_{\odot}$; the mass accretion rate is expressed as a fraction of the Eddington rate; and the radius is given in Schwarzschild radii, $2GM/c^{2}$.)  Nominally, then, the disk may not be truncated -- or, at least not restricted to very large radii -- in M81*.

Line spectra have not enabled much more clarity than the continuum.  Hubble UV spectroscopy revealed Ly$\alpha$ and C IV emission lines with FWHM$=3000-4000~{\rm km}~{\rm s}^{-1}$ \citep{ho1996}.  These lines may suggest a weak optical broad line region (BLR) -- potentially similar to the structures observed in Seyferts -- but they could also be interpreted as emission from the inner face of a truncated disk.  In X-rays, emission and absorption lines from the Fe K shell are the best probes of the inner accretion flow (see, e.g., \citealt{gmc2023}).  The width of the neutral Fe~K$_{\alpha}$ emission line revealed in Chandra gratings spectra limits the emitting region to ${\rm r} \geq 10^{3}~{\rm GM}/{\rm c}^{2}$ \citep{young2018}, assuming that the line is broadened by Keplerian motion (see \citealt{ishisaki1996}, \citealt{pellegrini2000}, \citealt{dewangan2004}, \citealt{page2004},    for additional detections and studies of the neutral Fe~K$_{\alpha}$ line).

Evidence of weak Fe XXV and Fe XXVI lines in stacked Chandra spectra could indicate emission from within a RIAF \citep{young2007}, or emission in a hot magnetohydrodynamic wind \citep{shi2021}.  These ionized emission lines were found to have a velocity width of ${\rm FWHM} \simeq 1500~{\rm km}~{\rm s}^{-1}$ -- slightly narrower than the UV and optical lines that are observed in M81*.  Interestingly, the Fe XXVI line is found to be red-shifted by approximately ${\rm v} \simeq 2600~{\rm km}~{\rm s}^{-1}$, nearly 0.01c \citep{young2007}.  

To better understand accretion onto massive black holes at low Eddington fractions, XRISM \citep{tashiro2024} observed M81* during its Performance Verification phase.  It was selected because its X-ray flux and line spectrum are likely to deliver the best possible view of a source class that represents a large fraction of the local AGN and galaxy populations.  Section 2 details the observation and the data reduction procedures.  The results of different analysis methods are described in Section 3.  In Section 4, we discuss the strengths and limitations of the present analysis, implications for LINERs and LLAGN, and opportunities for learning more.  Finally, we state a few conclusions in Section 5.

\section{Observations and Data Reduction}

M81* was observed by the X-ray Imaging and Spectroscopy Mission (XRISM) on 2024 May 8-12. XRISM is a JAXA/NASA collaborative mission, with ESA participation, and consists of two instruments: a high-resolution spectrometer (Resolve; \citealt{ishisaki2022}) and a soft X-ray imaging spectrometer (Xtend; \citealt{hayashida2018}).  The instruments sit behind identical X-ray Mirror Assemblies (XMA).  M81* was placed close to the center of the Resolve field of view.  The Resolve and Xtend instruments were operated in ``PX\_NORMAL'' model and ``full window'' mode, respectively.  

Prior to reduction and analysis, both the Resolve and Xtend data were reprocessed using the pre-release Build 8 XRISM software and calibration database libraries.  The rise-time screening was applied to the Resolve data, including the filtering out "STATUS[4]==b0" events. During the observation period of M81*, large solar flares occurred, and XRISM detected flare-related events.  The larger field of view in the Xtend instrument allowed the flux from M81* to be separated from background emission from the solar flares, and tracked within 2--10 keV light curves.  In this manner, we were able to effectively screen against flaring periods, achieving a net exposure of 179.8~ks.  Please see Appenix A for full details of this procedure.

The Resolve spectrum was created using only high-resolution primary (Hp) events, excluding pixel 27.  A redistribution matrix file (RMF) was created using the \texttt{rslmkrmf} task using the Hp events and the {\tt xa\textunderscore rsl\textunderscore rmfparam\textunderscore 20190101v006.fits} file.  Our RMF file includes the Gaussian core, exponential tail, Si K alpha emission lines, escape peaks, and the electron loss continuum.  The \texttt{xaarfgen} task was used to create a point source ancillary response file (ARF). 

A full analysis of the Xtend data is deferred to a subsequent paper.  Similarly, although we have filtered against solar flaring intervals, those intervals contain data that are useful for understanding the flares and the local space environment.  Investigations are currently underway, but are also deferred to a subsequent paper.

\section{Analysis and Results} 
The failure of the Resolve gate valve to open nominally truncates spectra below 1.6~keV.  Inspection of the data suggests the presence of instrumental features and/or flux calibration uncertainties in the lowest bins, so we elected to only fit the data above 1.7~keV.  At high energy, it is clear that the signal-to-noise ratio of the spectrum drops rapidly above 10~keV.  We set an upper bound of 10.7~keV to enable a 9~keV pass band for spectral fits.  The source shows negligible flux variability across the exposure, enabling a study of the time-averaged spectrum from the full observation.  In all of the work reported below, the reported errors are $1\sigma$ errors (68\% confidence).  

\subsection{Phenomenological Modeling of the Neutral and Ionized Fe~K Emission Lines}

\begin{figure}[t]
    \centering
     \includegraphics[width=0.9\columnwidth]{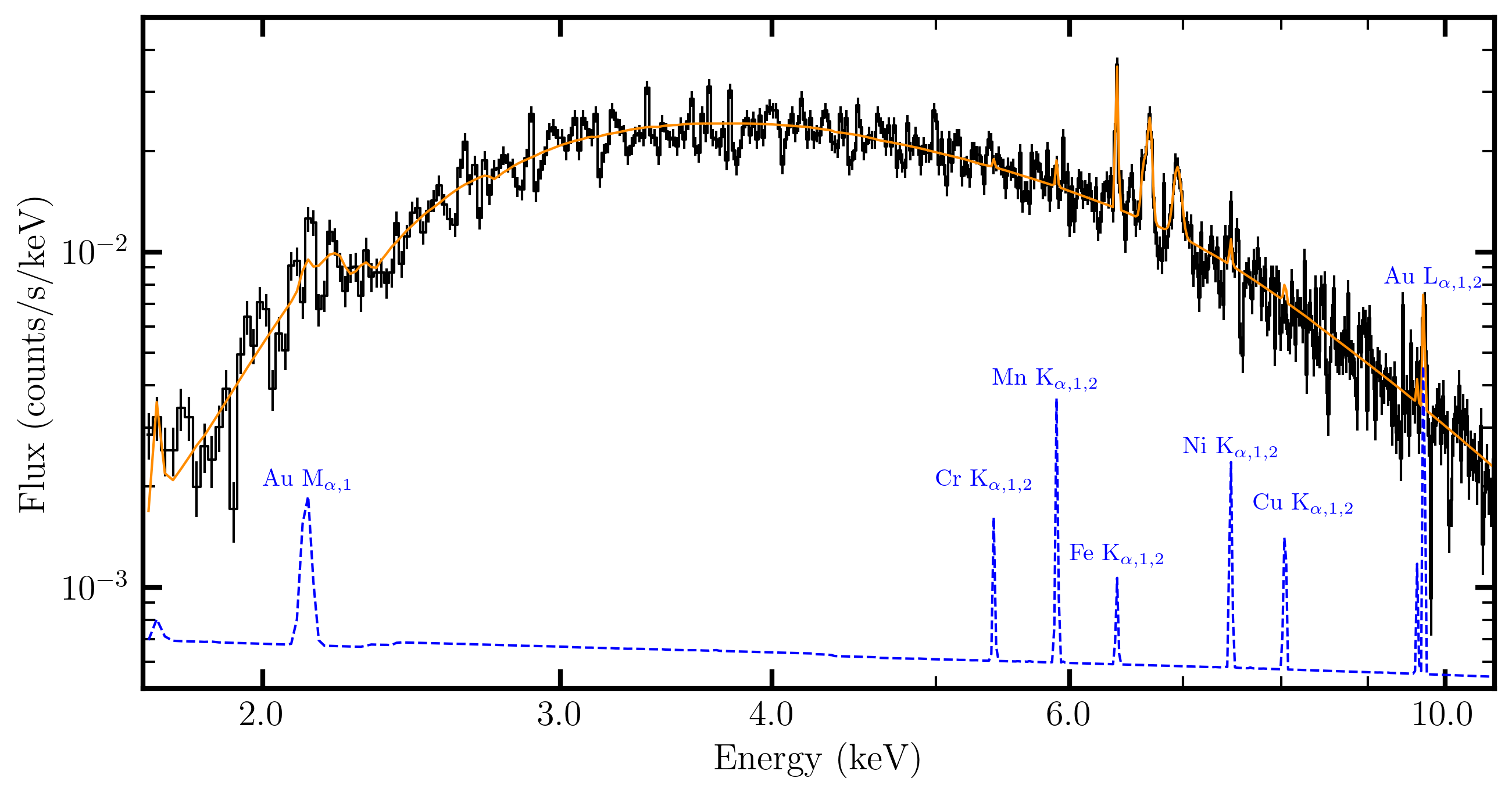}
     \includegraphics[width=0.9\columnwidth]{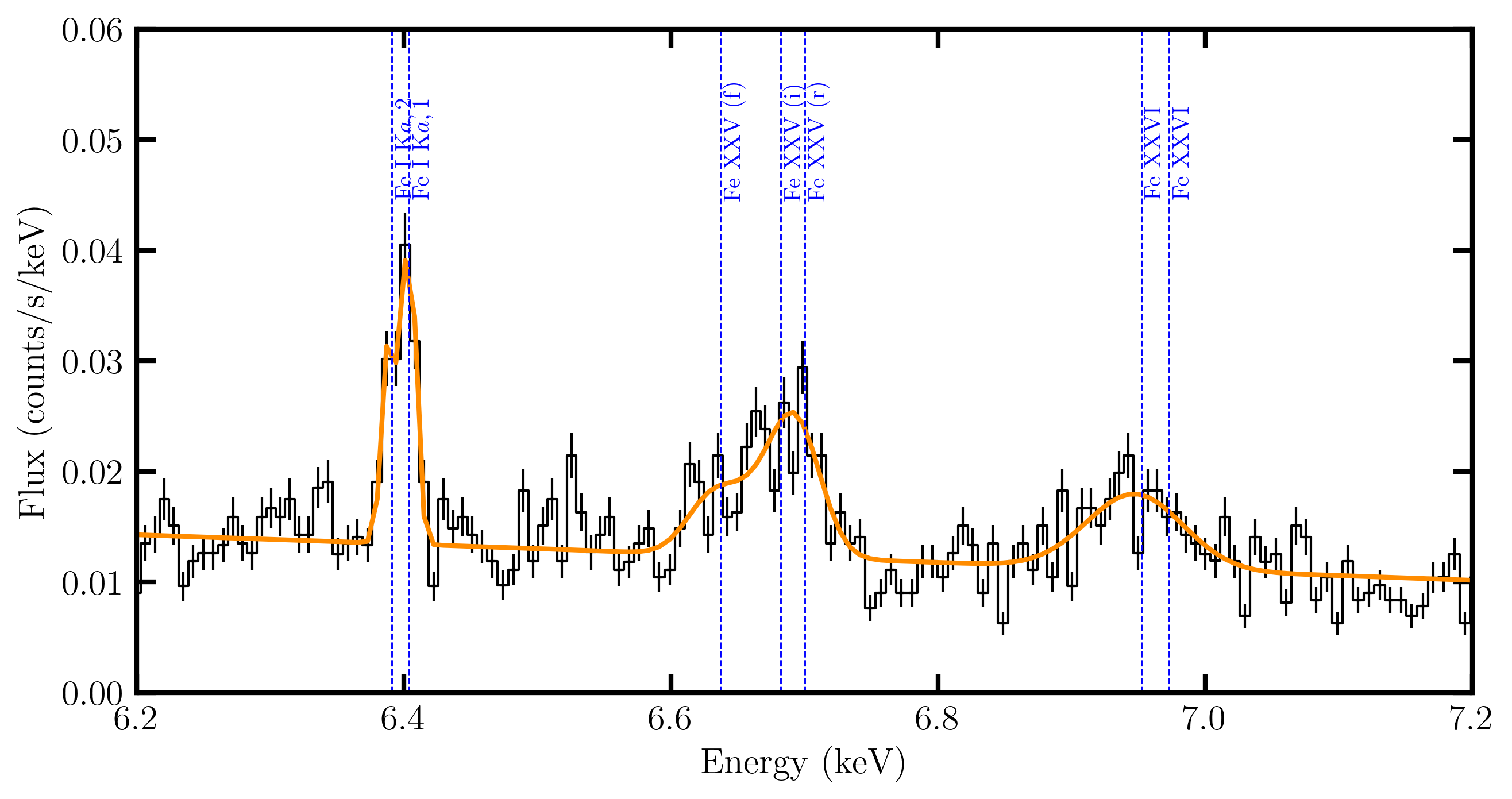}
   \caption{The Resolve spectrum of M81*.  Top: 
   The 1.7--10.7~keV band, binned using the ``optimal'' algorithm.  The fit shows a model consisting of a simple power-law and Gaussian functions in XSPEC (see Section 3.1).  The trend in blue shows the predicted non-X-ray background (NXB).  Prominent NXB lines are labeled.  The NXB is negligible in the band containing the Fe~K emission lines observed from M81*.  Bottom: The same spectrum and model, shown on a smaller pass band and on a linear scale.  Laboratory energies for lines observed from M81* are shown in blue.  Both neutral Fe~K$_{\alpha,1}$ and Fe~K$_{\alpha,2}$ lines are clearly detected, as well as the He-like Fe~XXV and H-like Fe XXVI complexes.  Please see Section 3.1 and Table 1 for details.}
   \label{fig:fig1}
\end{figure}

The Resolve spectrum of M81* obtained from May 8-12 is shown in Figure\,\ref{fig:fig1}.  Compared to NGC 4151 \citep{miller2024}, for instance, the flux observed from M81* is modest, and instrumental backgrounds may contribute to the observed spectrum.  Figure 1 shows the total spectrum observed with Resolve in black, and a synthetic non-X-ray background (NXB) spectrum in blue.  The NXB spectrum was generated using the XRISM NXB database.  No significant NXB lines are present in the Fe-K spectra region of interest (bottom panel in Figure\,\ref{fig:fig1}).

Since there is no consensus on how lines relate to specific accretion flow geometries within LLAGN, it is important to provide a model-indepenent description of the data.  Therefore, 
we began our analysis of the Resolve spectrum of M81* by considering a power-law continuum and simple Gaussian emission lines.  These fits were made using XSPEC version 12.14.1 \citep{arnaud1996}.  The resolve spectrum was binned using ftool~\texttt{ftgrouppha} with optimal binning \citep{kaastra2016}.  Then, the binned spectrum in the 1.7–10.7 keV band was fitted with the simple power-law and multiple-Gaussian model (via ``zgaussian'' components within XSPEC), yielding C-stat = 2328 (2282 d.o.f.). The best fitted parameters are listed in Table 1.

The bottom panel of Figure\,\ref{fig:fig1} zooms into the Fe-K region. The K$_{\alpha,1}$ and K$_{\alpha,2}$ emission lines are distinct.  The systemic redshift of M81* ($z=-0.000113$) was measured in optical light; for comparison, we fixed the energy of Fe K$_{\alpha,1}$ to 6403.8 eV and fitted the redshift as a free parameter. The best fit value of $z=0.0000\pm 0.0003$ is consistent with the optical systemic redshift.  The error on the redshift corresponds to an uncertainty of just $\sim$2 eV in the line centroid.  The line width was also determined to be $\sigma = 4.2^{+1.2}_{-0.8}$ eV, which corresponds to a velocity of about 460 km s$^{-1}$ (FWHM).  The line is therefore formally resolved.

We next set the redshift of the line components to the optical value, and measured the central energy and intensity of K$_{\alpha,2}$ as free parameters. The central energy of K$_{\alpha,2}$ is 14.8$^{+1.6}_{-1.8}$ eV smaller than K$_{\alpha,1}$, formally consistent with the expected energy difference in neutral iron of 13\,eV. 
However, the ratio of the {\em intensities} of the K$_{\alpha,1}$ and K$_{\alpha,2}$ components is 1.3$\pm$0.4, which is smaller than the expected value of 2.0.  This may signal that a weaker, broader, and potentially shifted neutral Fe~K$\alpha$ emission line is simultaneously present in the data, distorting the flux ratio of the narrow line components.

Young et al. (2007) pointed out a detection of a weak K$_{\beta}$ line of 3.1$^{+2.8}_{-2.7}\times10^{-6}~{\rm ph}~{\rm cm}^{-2}~{\rm s}^{-1}$ at E$\sim$7.057 keV. However, we obtained only an upper limit of 2$\times10^{-7}~{\rm ph}~{\rm cm}^{-2}~{\rm s}^{-1}$  at E$=$7.057 keV.  The K$_{\beta}$ line could plausibly be obscured by an absorber at a modest blue shift, but this represents a degree of fine-tuning beyond the scope of the present analysis.

\citet{page2004} fit spectra of M81* obtained with XMM-Newton, and measure an Fe~K$_{\alpha}$ line equivalent width of EW$=35\pm 13$~eV.  For their reported continuum, this corresponds to a line flux of $4.0\pm 1.3\times10^{-6}~{\rm ph}~{\rm cm}^{-2}~{\rm s}^{-1}$.  Thus, within errors, the narrow Fe~K$_{\alpha}$ line flux measured in prior observations is consistent with current flux levels; however, the large fractional errors may only mask variations in the line flux in response to variations in the ionizing flux.

In our simple fits to the Resolve data, the maximum energy difference between the K$_{\alpha,1}$ and K$_{\alpha,2}$ line components is 17~eV, within $1\sigma$ error bounds, so the allowed ``extra'' line splitting is $\Delta {\rm E} \leq 4$~eV.  Line splitting through the Zeeman effect causes splitting equal to $\Delta {\rm E} = 11.6~{\rm eV} ({\rm B}/10^{9})~{\rm Gauss}$.  Thus, the observed limit on the line splitting constrains the magnetic field in the emission region to be ${\rm B} \leq 3.5\times 10^{8}$~Gauss.

The data clearly reveal the He-like Fe XXV line between 6.6--6.7~keV. The line consists of four components, the resonance ($r$), two intercombination lines ($i$), and the forbidden ($f$) line.  Although velocity broadening makes the individual components somewhat indistinct, we fit the total line complex with four Gaussians at lab values, and with linked velocity shift and broadening parameters.  From this model, we are able to derive the $R$ = $f/i$ and $G=(i+f)/r$ ratios for simple plasma diagnostics. 

The $R$- and $G$- ratios are estimated to be $R = 1.4\pm1.0$ and $G = 0.7\pm0.3$, respectively. The $G$-ratio suggests that the origin of the plasma is collisional, but a photoionized plasma is allowed by the uncertainties. The line width was determined to be $19^{+4}_{-3}$~eV, which corresponds to  $\sim 2000$\,km\,s$^{-1}$ (FWHM). The redshift has large uncertainties and is consistent with the systemic redshift.

A broad line consistent with Ly$\alpha$ of H-like Fe XXVI is also detected in the data.  This line is a spin-orbit doublet, but turbulent broadening again appears to be even higher than for the He-like complex.  We therefore fixed the Gaussians to have the expected energy separation, enforced a 2:1 intensity ratio, and linked the velocity broadening of the Gaussians.  The line widths are measured to be  $35\pm 9$ eV, which corresponds to $\sim 3500$\,km\,s$^{-1}$ (FWHM). 
That line width is larger than  that of the broad H$\alpha$ line (2743\,km\,s$^{-1}$, \citealt{devereux2003}).  With this simple model, a significant red-shift is measured; which is discussed in more detail below using more physical models.

\subsection{The Neutral Fe~K Line: Physical Models}
Following the procedure used to model the narrow Fe~K$_{\alpha}$ line
complex in NGC 4151 \citep{miller2024}, we fit the line in the spectrum of M81 in SPEX \citep{kaastra1996} with a ``mytorus'' line function \citep{murphy2009}, modified by the ``Speith'' convolution model \citep{speith1995}.  ``Mytorus'' is fit as a table model; it describes the neutral Fe~K complex in terms of K$_{\alpha,1}$ and K$_{\alpha,2}$ lines, and
associated K$_{\beta}$ emission.  The model is also sensitive to the
column density of the emitting gas, by self-consistently including the
``Compton shoulder'' expected at 6.25~keV in the presence of cold
electrons.  The ``Speith'' model is an advanced convolution function
that accounts for the line shifts and enhancements expected across a
broad range of radii, assuming Keplerian orbits around a black hole.  Its parameters include the inner and outer emission radii ($r{1}$ and $r_{2}$, in units of $GM/c^{2}$), the emissivity index ($q$, as per $J\propto r^{-q}$), the inclination of the emitter, the spin of the black hole, as well as parameters that allow for a broken emissivity function and limb brightening adjustments (all held fixed at default values in our fits).

We fit the ``mytorus'' line function that assumes a power-law cut-off
energy of $E=$100~keV, and coupled the power-law index to that of the
directly observed power-law.  The other model parameters were also
allowed to vary, including the line normalization (defined relative to
the power-law normalization), the column density of the emitting gas,
the bulk shift of the emitting gas, and the inclination of the
emitting gas.  The inclination was coupled to the same parameter in
the ``Speith'' model for consistency.  After fixing an emissivity
profile index of $q=3$ (appropriate for a flat disk and an isotropic source)
and a black hole spin of $a = 0.7$, as per typical values in \citet{reynolds2021}, we fit for the inner radius of the line production region.  All other parameters within the ``Speith'' model were frozen at default values.  (We note that the spin value does not affect the radius measurements discussed below.)

\begin{figure}[t]
    \centering
     \includegraphics[width=0.9\columnwidth]{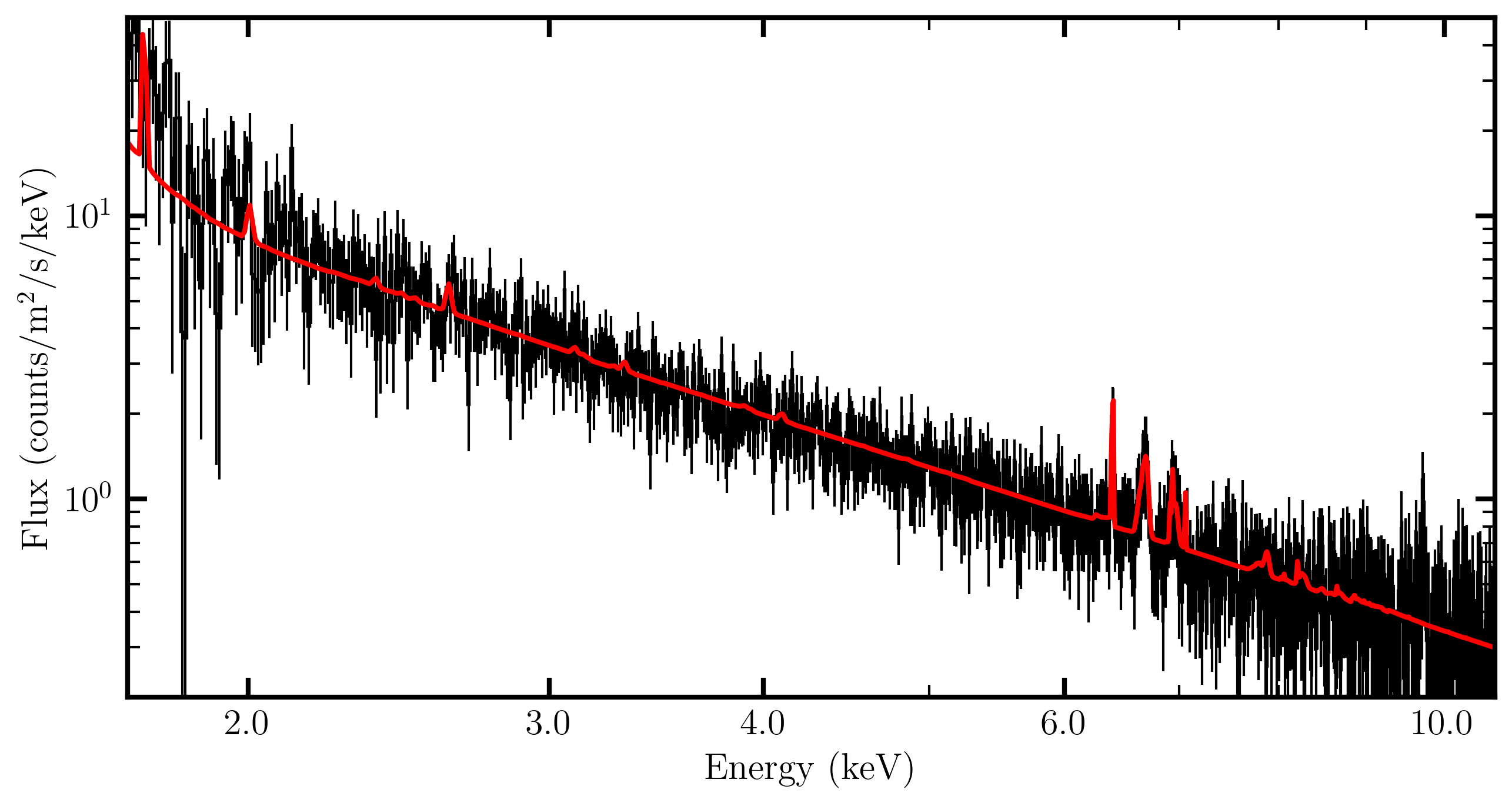}
     \includegraphics[width=0.9\columnwidth]{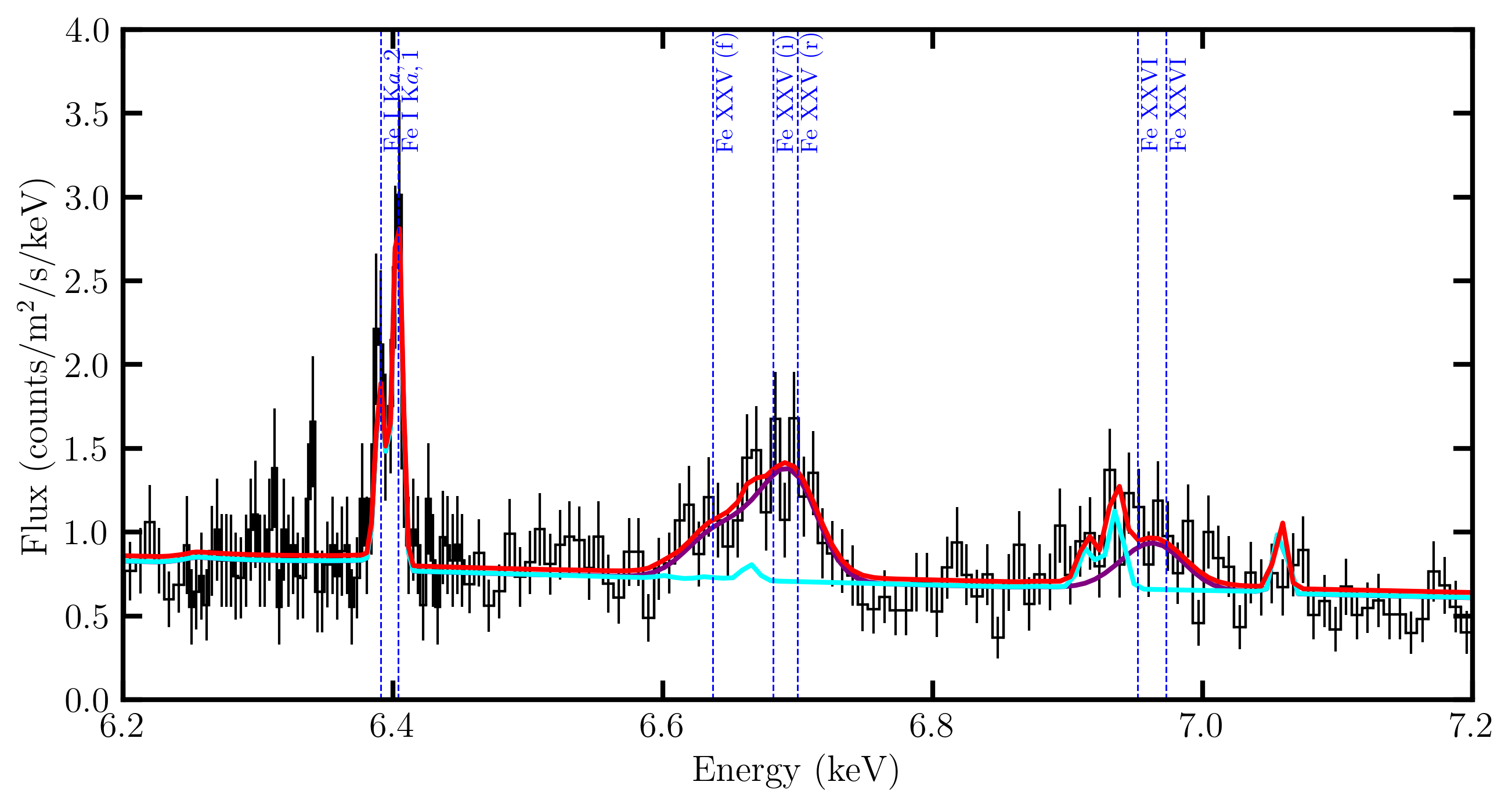}
   \caption{The Resolve spectrum of M81.  The data are binned using the ``optimal'' algorithm, and then by an extra factor of 2.0 except close to the Fe~K$_{\alpha}$ line.  Here, the ionized emission lines were fit in SPEX using the ``pion'' photoionization model.  Two zones of photoionized emission are required to fully describe the Fe~XXV and Fe~XXVI line complexes.  The contribution of the less ionized zone is shown in purple, while the more highly ionized and redshifted zone is shown in cyan.  Note that the statistical significance of the higher-ionization, redshifted zone is marginal. 
 Please see Sections 3.2 and 3.3, and Table 2 for details.}
   \label{fig:pion}
\end{figure}

\begin{figure}[t]
    \centering
     \includegraphics[width=0.9\columnwidth]{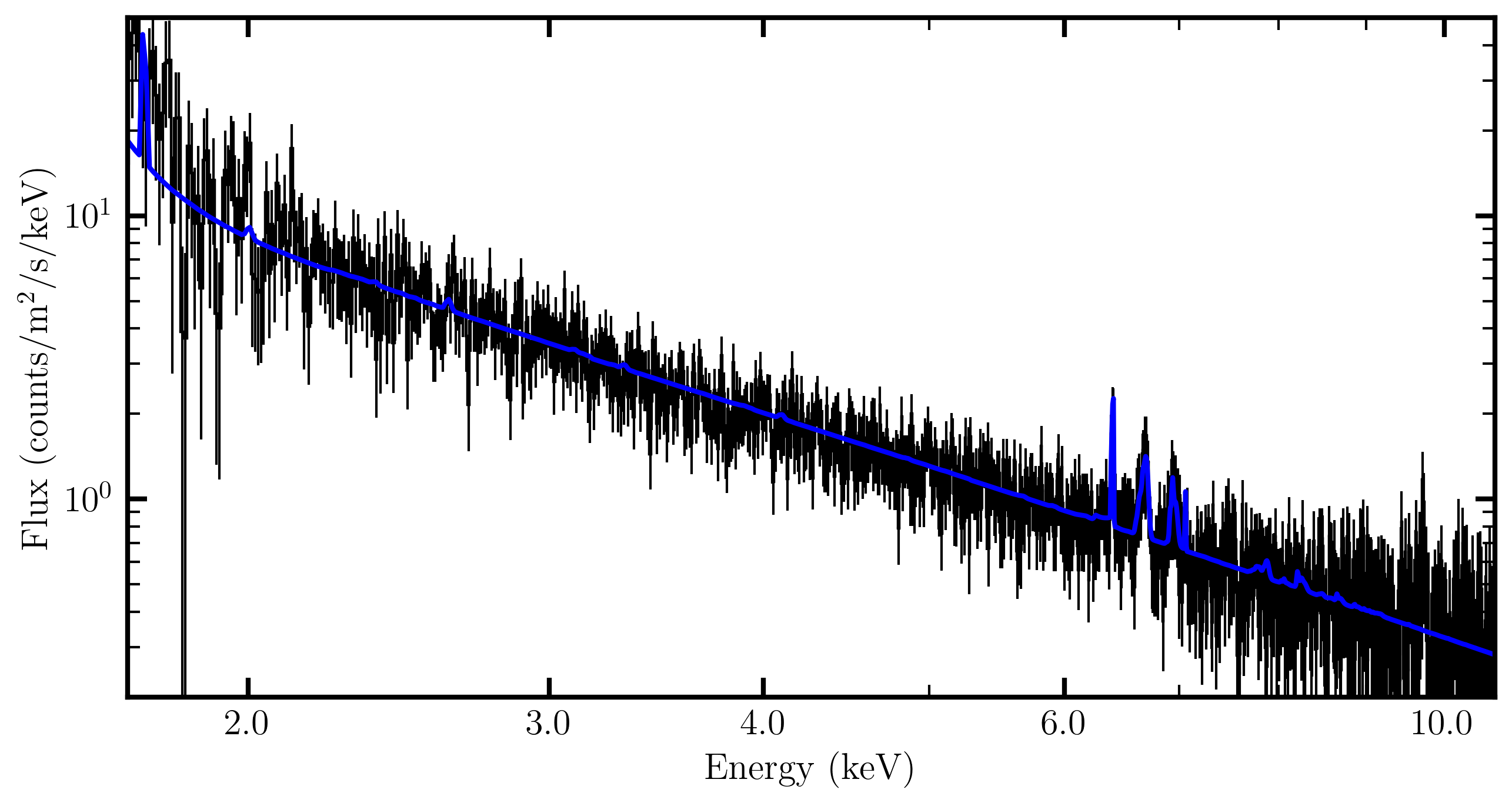}
     \includegraphics[width=0.9\columnwidth]{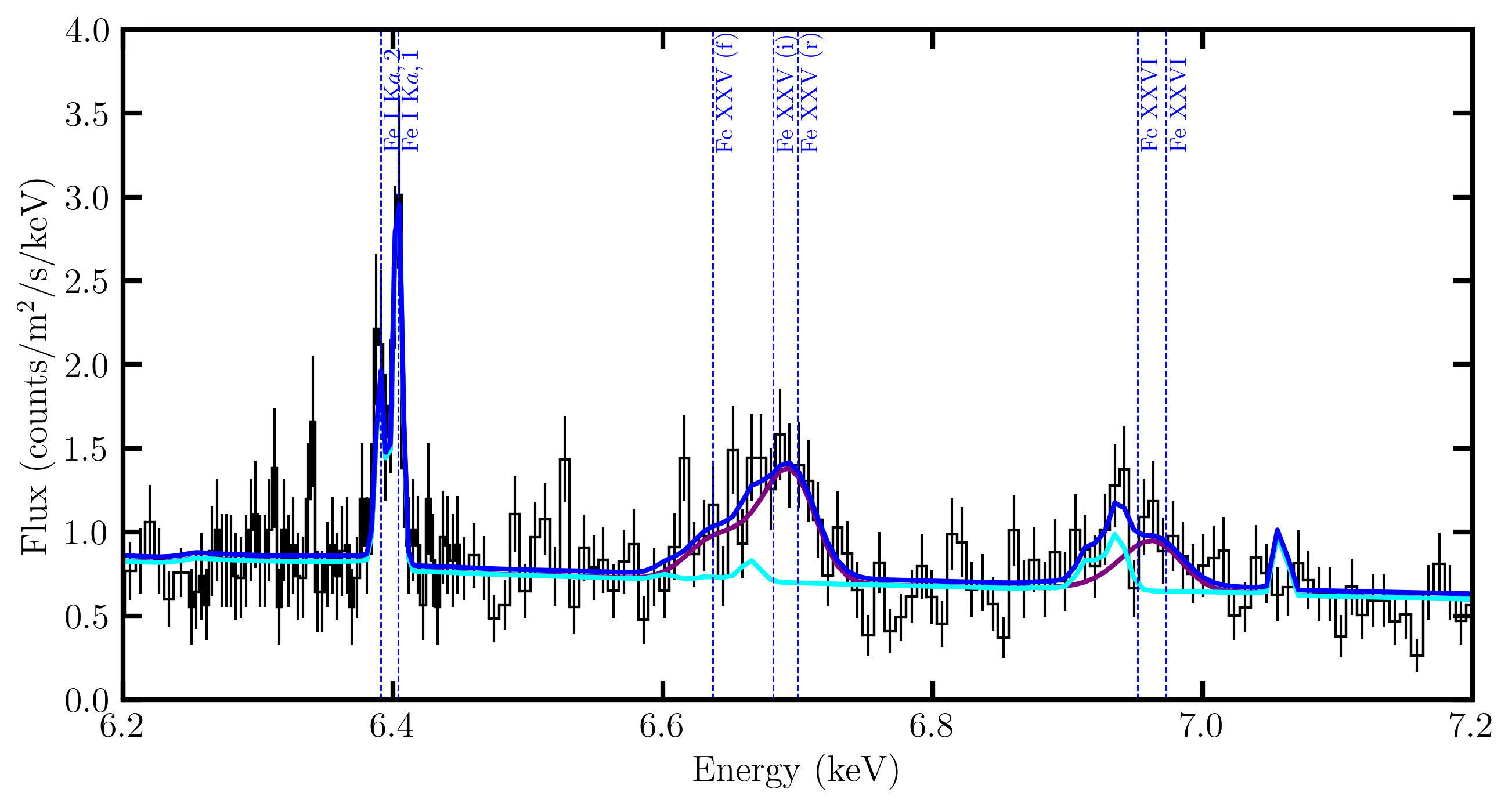}
   \caption{The Resolve spectrum of M81.  The data are binned using the ``optimal'' algorithm, and then by an extra factor of 2.0 except close to the Fe~K$_{\alpha}$ line.  Here, the ionized emission lines were fit in SPEX using the ``CIE'' collisional ionization model.  Two zones of collisional emission are required to fully describe the Fe~XXV and Fe~XXVI line complexes.  The contribution from the lower-temperature zone is shown in purple, while the contribution from the hotter zone is shown in cyan. Note that the statistical significance of the higher-temperature, redshifted zone is marginal. Please see Sections 3.2 and 3.4, and Table 3 for details.}
   \label{fig:cie}
\end{figure}

In initial fitting experiments, we found that the column density could not be constrained, so we fixed this parameter to the Compton-thick threshold appropriate for cold, dense gas, ${\rm N}_{\rm H} = 1.6\times 10^{24}~{\rm cm}^{-2}$.  When the other parameters are all allowed to vary freely, the constraints on the radius and inclination of the emission region are only weakly constrained.  In the best-fit photoionization model, for instance, the inner radius of the emitting region is measured to be ${\rm r} = 5^{+95}_{-4}\times 10^{4}~{\rm GM}/{\rm c}^{2}$, and the inclination is measured to be $\theta = 12^{+19}_{-12}$~degrees.   The range of inclinations encompasses constraints from modeling Hubble UV spectra of M81, $\theta = 14\pm 2$~deg \citep{devereux2003} (note, however, that jet-based constraints in the inclination are less restrictive; see \citealt{king2016}).  If the range of inclinations are restricted to those allowed in UV modeling, tighter and more consistent radius constraints are derived in our best fit photoionization and collisional ionization models: ${\rm r} = 6.3^{+9.9}_{-3.6}\times 10^{4}~{\rm GM}/{\rm c}^{2}$ and ${\rm r} = 4.1^{+11.5}_{-1.3}\times 10^{4}~{\rm GM}/{\rm c}^{2}$, respectively.  A conservative framing of these results is that the line region is constrained to a radius greater than ${\rm r} \geq 2.7\times 10^{4}~{\rm GM}/{\rm c}^{2}$.  These fits are detailed in Tables 2 and 3, and shown in Figures \ref{fig:pion} and \ref{fig:cie}. 

The equivalent width of the Fe~K$_{\alpha}$ line is EW$=37^{+7}_{-6}$~eV.  In the case of reflection from a neutral, optically thick slab, ${\rm R} \sim {\rm EW}/180~{\rm eV}$, where $R$ is the ``reflection fraction'' and normalized by $2\pi$ steradians.   Assuming that the observed line arises in these conditions, the measured equivalent width implies a reflection fraction of ${\rm R} = 0.21^{0.04}_{-0.04}$.  For contrast, a disk extending between the ISCO and infinity would subtend $2\pi~{\rm SR}$ and have a reflection fraction of ${\rm R}=1$, assuming an isotropic source of ionizing flux.  A truncated disk would have a much lower reflection fraction, and any BLR or torus with a vertical extent above a truncated disk might be expected to dominate the line flux.

\subsection{Photoionization Modeling of the Ionized Emission Lines}
A photoionization-based model for the 1.7--10.7~keV spectrum was constructed using the ``pion'' model within SPEX (\citealt{miller2015}, \citealt{mehdipour2016}).  The entire spectrum was binned using the ``optimal'' binning algorithm of \cite{kaastra2016}.  The continuum was fit assuming a simple power-law model, bent toward zero flux at low and high energy using two ``etau'' components.  This provides a power-law in the observed band but prevents an unphysical continuum outside of this band.  This incident spectrum was covered by two sequential ``pion'' layers.  We initially considered only photoionized emission.  In each layer, the photoionization parameter (${\rm log}\xi$ in erg\,s$^{-1}$cm), turbulent velocity ($\sigma$), bulk velocity shift ($v$), and covering factor ($\Omega$) were free parameters.  The gas column density is degenerate with the covering factor, so we fixed a value that is consistent with the data in each zone, ${\rm N}_{\rm H} = 5\times 10^{22}~{\rm cm}^{-2}$, and fit for the relative covering factor as this may be of more interest.  The hydrogen number density was fixed at a low value as no density-sensitive lines are evident in the spectrum.  

The results of this model are detailed in Table 2, and shown in Figure~\ref{fig:pion}.  Including fits to the neutral Fe~K$_{\alpha}$ line, the model achieves an excellent fit, with a Cash statistic of $C = 2420$ for $\nu = 2358$ degrees of freedom.  The power-law index is tightly constrained ($\Gamma = 1.89^{+0.08}_{-0.01}$), as is the normalization ($4.45^{+0.04}_{-0.05}\times 10^{4}$ in units of $10^{44}~{\rm ph}~{\rm s}^{-1}~{\rm keV}^{-1}$ at 1 keV), indicating a robust continuum determination.  The components within the Fe~XXV and Fe~XXVI complexes are mostly blended, indicating that the emission is dominated by gas with a large internal velocity shear. 
This is accounted for by the less-ionized zone, characterized by ${\rm log}\xi = 3.16\pm 0.09$; it has an internal 
velocity of $\sigma = 860^{+200}_{-170}~{\rm km}~{\rm s}^{-1}$, and a low bulk velocity of ${\rm v} = 160^{+230}_{-220}~{\rm km}~{\rm s}^{-1}$.  The more highly ionized zone, characterized by ${\rm log}\xi = 3.8^{+0.8}_{-0.5}$, primarily accounts for the narrow red-shifted flux in the Fe XXVI complex.  Accordingly, it carries a large bulk red-shift of ${\rm v} = 1600^{+150}_{-170}$\,km\,s$^{-1}$.  The measured covering factors are likely unphysical and only meaningful in a relative sense given that we fixed the same arbitrary value of ${\rm N}_{\rm H}$ for each zone.  The covering factor of the more highly ionized zone is poorly determined because its statistical significance is marginal, with the change in the Cash statistic similar to the change in the number of model parameters.

The ``mytorus'' model self-consistently predicts an Fe~K$_{\beta}$ line with a strength relative to the Fe~K$_{\alpha}$ line that is set by atomic physics.  This weak Fe~K$_{\beta}$ line is not evident in the data.  This might be explained if a photoionized absorption zone predicts a blue-shifted Fe XXVI line at the same energy.  This is notionally consistent with the appearance of putative weak absorption lines between the Fe XXV and Fe XXVI line complexes.  However, the data cannot require nor reject the presence of a blue-shifted wind that could describe some of these weak features.  Adding an absorption zone with the column density, ionization parameter, and velocity broadening fixed to emission zone 1 (see Table 2) suggests an absorption line with a blue-shift of ${\rm v} = -2700\pm 900~{\rm km}~{\rm s}^{-1}$, with a covering factor of $f_{c} = 0.33\pm 0.06$.  However, the Cash statistic is only reduced by $\Delta C = -8$ for an extra two degrees of freedom.   Adding an absorption component that is the counterpart of the more highly ionized emission zone does not improve the fit.  Similarly, there is no strong statistical support for additional emission zones that could potentially account for weak features between the Fe~K$_{\alpha}$ line and Fe~XXV complex.

\subsection{Collisional Modeling of the Ionized Emission Lines}

We also developed a SPEX model for the ionized Fe emission lines using collisional excitation components.  In this model, the two ``pion'' components were replaced with ``CIE'' components, describing a gas in collisional ionization equilibrium.  The CIE models do not include a bulk velocity shift, so each was acted upon by a ``reds'' component to introduce a variable red-shift.  Each CIE component includes many parameters, but many of them are beyond the reach of the modest spectrum observed from M81*.  We elected to fit for just three parameters from each CIE component: the flux normalization (the emission measure, or ${\rm Y}={\rm n}_{\rm e} {\rm n}_{\rm H} {\rm V}$, where ${\rm V}$ is the emitting volume), the electron temperature of the plasma, and the rms velocity broadening within the plasma.  (The abundances of all elements were frozen at solar values.)

This model is detailed in Table 3, and shown in Figure\,\ref{fig:cie}.  A good fit is also achieved with this model, with $C = 2429$ for $\nu = 2308$ degrees of freedom.  Similar to the results obtained using photoionization components, one CIE component is able to describe the bulk of the Fe~XXV and Fe~XXVI emission lines, while the second component is poorly constrained and describes weak but highly red-shifted flux in the Fe~XXVI line.  We found that the high temperature of the second component could not be well constrained, so we fixed a compatible value of ${\rm kT} = 20$~keV within the model.  This accurately captures the line flux while not adding continuum emission that would cause the power-law index to differ from the values inferred using photoionization and in previous observations with NuSTAR \citep{young2018}.

\section{Discussion}
We have analyzed a XRISM/Resolve observation of M81*, the first LLAGN to be studied with an X-ray calorimeter.  This AGN is the most proximal and brightest example of a source class that includes many of the massive black holes in the local universe.  The neutral Fe~K$_{\alpha}$ emission line is resolved into K$_{\alpha,1}$ and K$_{\alpha,2}$ components.   Modest broadening of this line points to a low inclination, and a relatively large production radius.  Complex He-like Fe~XXV and H-like Fe~XXVI emission lines are also observed; they are broad, and may require two components to be described in full.  Here, we examine different ways of associating the observed lines with geometries, explore how the results constrain prevalent models for the accretion flow in LLAGN, and comment on directions for future studies of M81* with XRISM.

\subsection{Comparisons to RIAF models}

If the narrow, neutral Fe~K$_{\alpha}$ line revealed with Resolve traces the inner edge of a thin accretion disk, then it offers a very strong confirmation of disk truncation at low Eddington fractions, as predicted by RIAF models \citep{quataert1999}.  Our results would then locate the inner accretion disk orders of magnitude farther from the black hole than prior estimates based on gratings and CCD-resolution spectra (\citealt{young2007}, \citealt{young2018}).  However, the narrow Fe~K$_{\alpha}$ line may not trace the inner disk.  Most probes of the central engine in LLAGN are really an extension of the spectroscopic tools that are utilized in Seyferts, where connections between lines and geometries are better established.  In Seyferts, however, the inner disk is likely traced by relativistically broadened Fe~K emission lines (e.g., \citealt{reynolds2021}), and the low flux of M81* simply does not reveal such features.

\citet{ho1996} report the detection of broad UV lines with FWHM$=2500~{\rm km}~{\rm s}^{-1}$ in Hubble/STIS spectra, generally consistent with values in Seyfert AGN.  \citet{ho2008} interpret such lines as emission from BLR clouds, irradiated by an inner RIAF.  With an inferred radius of ${\rm r} \geq 2.7\times 10^{4}~{\rm GM}/{\rm c}^{2}$, the Fe~K$_{\alpha}$ line could originate from a BLR geometry, but this would represent the outer extent of a typical BLR.  Moreover, the broadening observed in the Fe~K$_{\alpha}$ line is only FWHM$=460~{\rm km}~{\rm s}^{-1}$; this is significantly narrower than the broad lines resolved with Hubble, and thus locate the Fe~K$_{\alpha}$ line at larger radii.  The H$\alpha$ line {\em core} has a velocity width of FWHM$=400~{\rm km}~{\rm s}^{-1}$ \citep{devereux2003}, and may trace the same geometry.

Resolve spectroscopy of the Seyfert-1.5 galaxy NGC 4151 reveals a narrow component in the Fe~K$_{\alpha}$ line that traces radii of ${\rm r} = 0.5-2\times 10^{4}~{\rm GM}/{\rm c}^{2}$, in close agreement with dust reverberation mapping of the inner wall of the torus \citep{miller2024}.  This is only just below the smallest radii that we infer in M81*.  Especially in view of these similarities, the simplest and most robust interpretation of the narrow Fe~K$_{\alpha}$ line in M81* is likely that it traces the inner wall of a remnant torus with a low covering factor, potentially tied to a waning $\dot{m}$ and AGN activity period.

This inference can be reconciled with prior data and results.  Joint fits to Suzaku and NuSTAR spectra of M81* did not find evidence of reflection, with an upper limit of ${\rm R}\leq 0.1$, and this would nominally rule out associating the narrow Fe~K$_{\alpha}$ line with a torus \citep{young2018}.  We reduced the same NuSTAR data, and fit it with the ``pexmon'' reflection model \citep{nandra2007}.  We obtain a reflection fraction of R$ = 0.14\pm 0.03$.  Fits with ${\rm R} = 0$ increase a Cash statistic by $\Delta C = 43$ for one degree of freedom, implying that reflection is weak, but significant at the $6\sigma$ level of confidence (Miller et al.\ 2025, in prep.).  The reflection fraction formally agrees with the value of ${\rm R} = 0.21\pm 0.04$ inferred from the equivalent width of the Fe~K$_{\alpha}$ line in the Resolve data.  We suggest that the inference of zero reflection in joint fits was driven by cross-calibration uncertainties between Suzaku and NuSTAR; these are alleviated when NuSTAR is considered alone.
The main thrust of \cite{young2018} is that reflection in M81* is much lower than in higher-Eddington AGN, signaling an altered accretion flow geometry; this outcome is unaffected if the reflection is small rather than zero.

Associating the bulk of the narrow Fe~K$_{\alpha}$ line observed in M81 with a remnant torus does not rule out contributions from a BLR.  Fits to the Fe~K$_{\alpha}$ line components with simple Gaussians did not quite register the 2:1 flux ratio that is expected.  This can potentially be explained if a weak BLR component to the line flux contributes in a manner that distorts the ratio of the torus component.  A future deep exposure with Resolve may be able to reveal a second component, and confirm a BLR geometry in addition to a torus.

\subsection{Ionized accretion flows}
Within radiatively inefficient flows, collisional emission and Compton upscattering are the dominant radiation mechanisms.  Although the inner parts of these regions are likely to be very hot and the gas is likely to be fully ionized, the outer extremes of a RIAF may be cool enough for Fe~XXV and Fe~XXVI to survive.  If the gas is collisionally ionized, it is possible that the zones detailed in Table 3 may describe part of the gas accreting in the RIAF zone.   Both CIE zones are redshifted, and the hotter zone has a much higher redshift ($1500\pm 200~{\rm km}~{\rm s}^{-1}$ versus 
$180^{+240}_{-60}~{\rm km}~{\rm s}^{-1}$).  The higher velocity corresponds to free-fall at ${\rm r} \sim 8\times 10^{4}~{\rm GM}/{\rm c}^{2}$.  Smaller radii would result after correcting for projection effects, and might be fully compatible with the truncation radius that is inferred if the inner disk is traced by the neutral Fe~K$_{\alpha}$ emission line.  The small velocity broadening seen in this component, $\sigma = 230\pm 230~{\rm km}~{\rm s}^{-1}$, is inconsistent with this radius, but could potentially reflect a sub-Keplerian flow with a sizable radial velocity component.  On the other hand, the cooler component has a very modest redshift, likely inconsistent with even a highly truncated disk.  Both zones may be incompatible with a disk that is truncated interior to the radii traced by broader UV lines.  For these reasons, it is unlikely (but not impossible) that the Fe~XXV and Fe~XXVI lines trace collisional gas within a RIAF.

Instead, the Fe~XXV and Fe~XXVI lines may trace a RIAF wind.  The outer extremes of these regions may be so hot that they are unbound, leading to a wind (\citealt{bb99}, \citealt{narayan99}).  In this case, the broad and redshifted emission that is observed must originate on the far side of the central engine.  A largely equatorial wind in a system that is viewed at a low inclination angle could give rise to a wind signature like this, and Hubble spectra suggest an inclination of just $\theta = 14\pm 2$~deg. in M81* \citep{devereux2003}.  In this case, the bulk velocities and broadening of the components in Tables 2 and 3 would depend strongly on the details of the wind: whether it is launched ballistically or with the aid of magnetic pressure, whether it diverges or holds a fixed solid angle.

The lines could also originate in a magnetohydrodynamic (MHD) wind, driven from a RIAF.  \citet{shi2021} developed an MHD wind model for M81* based on Chandra gratings spectra.  Translating the line fluxes reported in that work to the units used in Table 1, \cite{shi2021} measure Fe XXV and Fe XXVI line fluxes of $F = 5\pm 2\times 10^{-6}~{\rm ph}~{\rm cm}^{-1}~{\rm s}^{-1}$ and $F = 7^{+3}_{-2}\times 10^{-6}~{\rm ph}~{\rm cm}^{-1}~{\rm s}^{-1}$ in the more prominent red-shifted wind component.  These values are broadly compatible with the integrated line fluxes that we have measured using narrow line components that are suited to the calorimeter data.  A key feature of the MHD model is that it also predicts blue-shifted emission; the prior Chandra data do not offer a clear detection, giving a line flux of $F = 2.5\pm 2.0\times 10^{-6}~{\rm ph}~{\rm cm}^{-2}~{\rm s}^{-1}$.  Fitting a Gaussian at 7.046~keV as per \cite{shi2021}, and assuming a line width equal to the Fe XXVI line in the red-shifted wind component, we measure a best-fit line flux of just $F = 0.08\times 10^{-6}~{\rm ph}~{\rm cm}^{-2}~{\rm s}^{-1}$, and a 90\% confidence limit of $F = 0.5\times 10^{-6}~{\rm ph}~{\rm cm}^{-2}~{\rm s}^{-1}$.  On balance, it is likely that the predicted blue-shifted wind component is absent.  Additional observations could determine that the wind is variable, and refinements to the model may be able to accommodate these limits.

\subsection{Natal Fermi bubbles?}

Although Sgr~A* now accretes at an Eddington fraction that is orders of magnitude lower than that inferred in M81*, X-ray reflection nebulae in the nucleus of the Milky Way suggest that it may have been comparably luminous in the past (e.g., \citealt{ponti2010}).   Such episodes may be responsible for the extended bi-polar bubbles that are linked to Sgr A*, and clearly detected in $\gamma$-rays and X-rays \citep{su2010}.   Could the ionized emission in M81* be the natal precursor to large galactic bubbles?

A quick examination of Chandra imaging observations of M81* confirms that Fe~XXV and Fe XXVI emission is not extended, but rather confined to the nuclear source and consistent with the point spread function.  Conservatively adopting a value of 1 arc second for that radius, a spherical emitting geometry, and assuming a distance of ${\rm d} = 3.61^{+0.21}_{-0.19}$\,Mpc \citep{tully2016}, the collisional emission measures
detailed in Table 3 imply number densities of ${\rm n} \sim 15~{\rm cm}^{-3}$ and ${\rm n} \sim 10~{\rm cm}^{-3}$.  Estimating the total energy in that region via E$ = \frac{3}{2}{\rm nkTV}$, the total energy in each component is then ${\rm E}_1 \sim 1.7\times 10^{53}$~erg and ${\rm E}_2 \sim 3.1\times 10^{53}$~erg.  Dividing these by the characteristic luminosity in each component, ${\rm L}_1 \sim 3.2\times 10^{39}~{\rm erg}~{\rm s}^{-1}$ and ${\rm L}_{2} \sim 2.0\times 10^{39}~{\rm erg}~{\rm s}^{-1}$, gives characteristic radiative cooling times of ${\rm t}_1 = 1.7~{\rm Myr}$ and  ${\rm t}_2 = 4.9~{\rm Myr}$.  If the gas retains its velocity as it expands, the hotter component could travel ${\rm d} \sim 7$~kpc in that characteristic time.  Of course, the gas is likely to cool as it expands, and actual expansion would depend on buoyancy and drag forces.  Drag could be counterbalanced by a degree of continuous energy injection from ongoing shocks in the outflow.  Thus, it appears possible that the ionized gas could represent an early phase of bubble formation.

\subsection{Comparisons to MAD models}
Given that M81* launches jets, and MAD models are designed to explain jet production, it is important to examine the Resolve data in this context.  
Via fits to the components of the neutral Fe~K$_{\alpha}$ line, we find an upper limit of ${\rm B} \leq 3.5\times 10^{8}~{\rm Gauss}$ at a radius of ${\rm r} \geq 2.7\times 10^{4}~{\rm GM}/{\rm c}^{2}$.  Most MAD formulations instead work in terms of the magnetic flux that threads the black hole, $\Phi_{\rm BH} = \phi_{\rm BH}(\dot{M}_{\rm BH} {\rm R}_{\rm g}^{2}c)^{1/2}$, where $\phi_{\rm BH}$ is the dimensionless magnetic flux, $\dot{M}_{\rm BH}$ is the mass accretion rate onto the black hole, and ${\rm R}_{\rm g} = {\rm GM}_{\rm BH}/{\rm c}^{2}$.  In simulated MAD disks that launch jets, $\phi_{\rm BH} \simeq 30$, of course depending on assumed accretion rates (see, e.g., \cite{mckinney2012}, \cite{liska2022}).  The extra line splitting due to the Zeeman effect can be written as $\Delta {\rm E} = 27$~eV $(\phi/30) (m/10)^{-1/2} (\dot{m}/0.3)^{1/2} {\rm r}^{\rm -p}$ \citep{inoue2023} (where $r$ is in units of $GM/c^{2}$), and assuming the quantity ${\rm BR}^{\rm p}$ is constant as per a magnetic field that is frozen-in (here ${\rm p}$ is just a dimensionless index), and characterizing the mass and mass accretion rate as $m = M/M_{\odot}$ and $\dot{m} = \dot{M}/\dot{M}_{Edd}$.  

It is useful to consider two idealized cases: ${\rm p}=0$ corresponds to the MAD condition, where the magnetic flux is saturated, and ${\rm p}=2$, which corresponds to a disk wherein the magnetic field is conserved.  Assuming ${\rm M}_{\rm BH} = 7\times 10^{7}~{\rm M}_{\odot}$ \citep{devereux2003}, utilizing our estimates of $\dot{\rm M}/\dot{\rm M}_{\rm Edd} \simeq 10^{-5}$ and ${\rm r} = 2.7\times 10^{4}~{\rm GM}/{\rm c}^{2}$, the upper limit of $\Delta {\rm E} \leq 4$~eV on the splitting of the neutral Fe~K$\alpha$ components gives $\phi \leq 2\times 10^{5}$ for $p=0$, and far larger values for $p=2$.  Thus, even the most conservative case only gives a very weak limit, though one that allows for a MAD flow close to the black hole.  

Our results suggest that X-ray spectroscopy may not be an incisive probe of MAD models through line splitting, save in extremely fortuitous circumstances.  Pushing to higher mass accretion rates and achieving much tigher constraints on line splitting will help to set tighter limits on magnetic fields and flux.  In the abstract, these factors could be realized jointly in brighter sources since the higher mass accretion rate may tend to create a stronger Fe~K$_{\alpha}$ emission line.  However, in such AGN, the line is likely to be far more complex: narrow line components may reflect distant conditions more than the nature of the central flow, and lines associated with the central flow may be velocity broadened to a point where additional line splitting is difficult or impossible to measure (e.g., \citealt{miller2024}).   Moreover, although many MAD simulations extend out to ${\rm r} = 10^{5}~{\rm GM}/{\rm c}^{2}$, and a parameter space of MAD solutions may describe many facets of Sgr A* \citep{eht22}, most work is concentrated at higher mass accretion rates, and may not be fully applicable to LLAGN.  

\subsection{Outstanding questions}

Despite the advances made possible by Resolve, a number of outstanding questions remain.  First -- and perhaps most importantly -- the radial extent of the innermost accretion disk is still uncertain.  While it seems unlikely that a standard thin disk remains close to the black hole at ${\rm L}/{\rm L}_{\rm Edd} = 10^{-5}$, the revised mass of M81* may not require truncation \citep{quataert1999}.  A disk that bears some resemblance to those in Seyferts must remain at ${\rm r} \sim 10^{3}~{\rm GM}/{\rm c}^{2}$ in order to provide a platform from which BLR clouds can be launched.  Deep observations with XRISM may be able to detect contributions to the Fe~K$_{\alpha}$ line from the BLR.  

We have pursued fits the neutral Fe~K$_{\alpha}$ emission line that assume an emissivity of $q=3$ (defining the emissivity law as $J\propto r^{-q}$), which is appropriate for a flat disk.  If the BLR and/or torus actually represent a step function in vertical height with radius -- nearly a wall -- then an emissivity of $q=2$ would be more appropriate.  The XRISM data are unable to distinguish even the stark contrast posited here: when an emissivity of $q=2$ is instead assumed, fully consistent radius constraints are derived.  In practice, the BLR and torus may represent deviations from a flat disk, but with an opening angle that would require an intermediate emissivity index.  It is possible that the opening angle of such geometries depends on $\lambda_{Edd.}$.  Future observations of M81* and Seyferts may be able to explore the details of these geometries, and make key comparisons.

Although a wind is a plausible explanation of the ionized and redshifted emission that is observed with Resolve, the mechanism that produces this emission remains uncertain.  Variability in the line properties could lift this degeneracy; however, M81* did not appear to vary significantly during this observation, nor during overlapping {\it Swift} monitoring.  We attempted to detect line flux variability by dividing the observation into two segments, but could detect none, given the limited count rate.  On longer timescales, the soft X-ray {\it continuum} of M81* is known to vary, showing extreme harder-when-brighter behavior \citep{Connolly2016, Peretz2018}.  Fits to the Fe~XXV lines with simple models, photoionization models, and collisional excitation models all yield line widths that are broadly consistent with the FWHM$=1500~{\rm km}~{\rm s}^{-1}$ O VIII line observed with Chandra \citep{young2007}, likely signaling that future observations with the gate valve open could enable broader fits that reveal more about the plasma.  
However, it should also be noted that in Mrk~1239, for instance, the ionized emission lines are found to arise through a combination of photoionization and collisional excitation \citep{buhariwalla2023}.

\section{Conclusions}
Observations of the LLAGN M81* with Resolve have revealed a relatively narrow neutral Fe~K$_{\alpha}$ line.  In combination with prior UV studies, this suggests that diminished or fading Seyfert geometries likely persist at ${\rm L}/{\rm L}_{\rm Edd} = 10^{-5}$, even if the nature of the inner flow is a RIAF or MAD flow rather than a standard thin disk and corona.  Broad and red-shifted emission lines from He-like Fe~XXV and H-like Fe~XXVI are consistent with a wind, but potentially also consistent with a natal Fermi bubble geometry.  Future XRISM studies that focus on detecting variability in the neutral and ionized emission lines can help to determine the nature of the accretion flow geometry in this paradigmatic source.\\

We thank the referee for a careful review of this paper and numerous helpful suggestions that improved it.  JMM acknowledges the inspirational work ethic of Ivy E. Miller in pushing this work to a conclusion, and helpful conversations with Luis Ho, Yoshiyuki Inoue, and Eliot Quataert.  The team thanks helpful comments from Erin Kara.  JMM acknowledges funding from NASA through grant 80NSSC24K0369.  EB acknowledges the hospitality of the MIT Kavli Institute during the early phases of this work, and the support of NASA grant 80NSSC20K0733.

\clearpage

\appendix
\section{reduction of the solar-flare related events}
As described in section 2, the observation data was taken during the intense space storm from May 8th to 12th and, hence, can contain a significant number of solar-flare-related events above the nominal background ones. These additional contaminants can be roughly categorized into three components: the albedo K-alpha line from the neutral Oxygen in the Earth’s atmosphere, the particle-induced events, and the scattered solar X-ray emissions. The first component is a line feature at 0.53 keV, outside the effective band of Resolve with the gate-valve-closed configuration. However, the rest exhibit a continuum above 2 keV and the He-like iron K-alpha line at 6.7 keV, which may contaminate the Resolve spectrum. To mitigate the possible risk of contamination, we utilized the $2\--10$ keV Xtend light curve extracted from the entire field of view with apparent X-ray sources excluded. The wide field of view of Xtend ($30’ \times 30’$) enables us to track the timing of solar-flare-related events with high statistics. We generated a series of good time intervals that exclude all time regions wherein the Xtend 2-10 keV count rate exceeds 1.6 count/sec. The threshold count rate is defined as the mean$+2\sigma$ of the Gaussian that fits around the peak ($0.6\--1.3$ count/sec) of the $2\--10$ keV count rate distribution, i.e., the nominal background level. Although applying the good time interval discards $\sim23\%$ of the total exposure, the effect of the two solar-flare components drastically reduces to a nearly negligible level. 

\bibliography{ms}{}
\bibliographystyle{aasjournal}

\clearpage

\smallskip
\begin{deluxetable}{lcc}
\tabletypesize{\scriptsize}
\tablewidth{0pt}
\tablecaption{Best-fit Simple model
}
\label{tab:col}
\tablehead{
\colhead{Parameter}
& \colhead{Full Obs} 
& \colhead{Notes}
}
\startdata
PL slope $\Gamma$ & $2.01\pm0.02$ &  \\
PL norm [10$^{-3}$] & $3.102\pm0.005$  & \\
ph\,s$^{-1}$cm$^{-2}$keV$^{-1}$@1\,keV& &\\
&&\\
\hline
Fe K (cold) & & \\
$\alpha_1$ redshift & $0.0000\pm0.0002$
& frozen at 6.4038 keV\\
~~~~~$\sigma$ [eV] & $4.2^{+1.2}_{-0.8}$ & $\sim 460$\,km\,s$^{-1}$ (FWHM)\\
~~~~~intensity [10$^{-6}$] & $2.1\pm0.4$ 
& EW$\sim$24 [eV]\\
$\alpha_2$ center energy [eV] & $6389.0^{+1.6}_{-1.8}$  & redshift and $\sigma$ tied to K$\alpha$1\\
~~~~~intensity [10$^{-6}$] & $1.4\pm0.3$
& EW$\sim$16 [eV]\\
$\beta$ ~~~~~intensity [10$^{-6}$] & $< 0.19$
& EW$<$3 [eV], redshift and $\sigma$ tied to K$\alpha$1
, frozen at 7.0580 keV\\
&&\\
\hline
Fe XXV (He-like)& & \\
(r) redshift  & $0.0011\pm0.0007$ &  frozen at 6.7000 keV \\
~~~~~$\sigma$ [eV] & $19^{+4}_{-3}$ &$\sim 2000$\,km\,s$^{-1}$ (FWHM)\\
~~~~~intensity [10$^{-6}$] & $3.4\pm0.6$
& EW$\sim$41 [eV]\\\
(i) intensity [10$^{-6}$] & $ 1.0^{+0.6}_{-0.5}$
& EW$\sim$12 [eV]\\
(f) intensity [10$^{-6}$] & $ 1.4\pm0.5$
&  EW$\sim$16 [eV]\\
&&\\
\hline
Fe XXVI (H-like)& & \\
K$\alpha_1$~ redshift   & 0.0026$^{+0.0012}_{-0.0013}$ &  frozen at 6.972 keV \\
~~~~~~~~$\sigma$ [eV] & $35\pm9$ &$\sim 3600$\,km\,s$^{-1}$ (FWHM)\\
~~~~~~~~intensity [10$^{-6}$] & $2.5\pm0.5$
& EW$\sim$51 [eV]\\
K$\alpha_2$~~intensity [10$^{-6}$] & =0.5$\times$ K$\alpha_1$ & \\
&&\\
\hline
Cstat &2329& \\
d.o.f. &2282&  \\
\hline
\enddata
\end{deluxetable}

\clearpage

\smallskip
\begin{deluxetable}{lccc}
\tabletypesize{\scriptsize}
\tablewidth{0pt}
\tablecaption{Fe~K$_{\alpha}$ Line and Photoionized Emission Model
}
\label{tab:col}
\tablehead{
\colhead{Parameter} 
& \colhead{Value}  
& \colhead{Notes}  
}
\startdata
PL slope $\Gamma_1$ & $1.89^{+0.08}_{-0.05}$ & the slope below the break \\
PL norm [$10^{4}$] & $4.45^{+0.04}_{-0.05}$ & -- \\
\hline
Fe~K$_{\alpha}$ ${\rm N}_{\rm H}$ [$10^{24}~{\rm cm}^{-2}$] & $1.6$ & fixed \\
Fe~K$_{\alpha}$ norm. [$10^{4}$] & $3.31^{+0.04}_{-0.04}$ &  -- \\
Fe~K$_{\alpha}$ radius [$10^{4}~{\rm GM}/{\rm c}^{2}$] & $5^{+95}_{-4}$ & $6.3^{+9.9}_{-3.6}$ for $i=14^{+2}_{-2}$~deg \\
Fe~K$_{\alpha}$ inclination [degrees] & $12^{+19}_{-12}$ & -- \\
\hline
PI emis. 1 ${\rm N}_{\rm H}$ [$10^{22}~{\rm cm}^{-2}$] & $5.0$ & frozen \\
PI emis. 1 log~$\xi$ & $3.16^{+0.09}_{-0.09}$ & -- \\
PI emis. 1 $\sigma$ [${\rm km}~{\rm s}^{-1}$] & $860^{+200}_{-170}$ & -- \\
PI emis. 1 $v$ [${\rm km}~{\rm s}^{-1}$] & $170^{+230}_{-220}$ & consistent with zero \\
PI emis. 1 $f_{cov}$ & $1.5^{+0.4}_{-0.3}$ & -- \\
\hline
PI emis. 2 ${\rm N}_{\rm H}$ [$10^{22}~{\rm cm}^{-2}$] & $5.0$ & frozen \\
PI emis. 2 log~$\xi$ & $3.8^{+0.8}_{-0.5}$ & -- \\
PI emis. 2 $\sigma$ [${\rm km}~{\rm s}^{-1}$] & $210^{+130}_{-160}$ & -- \\
PI emis. 2 v [${\rm km}~{\rm s}^{-1}$] & $1610^{+150}_{-170}$ & --\\
PI emis. 2 $f_{cov}$ & $2.0^{+10.0}_{-1.8}$ & poorly constrained \\
\hline
Flux ($10^{-11}~{\rm erg}~{\rm cm}^{-2}~{\rm s}^{-1}$)  & $2.6\pm 0.1$ & 0.0136--13.6~keV \\
Lum. ($10^{40}~{\rm erg}~{\rm s}^{-1}$) & $4.6\pm 0.1$  & 0.0136--13.6~keV \\
\hline
Cstat/$\nu$ & 2420/2358 & 1.7--10.7~keV \\
\enddata
\end{deluxetable}

\clearpage

\smallskip
\begin{deluxetable}{lccc}
\tabletypesize{\scriptsize}
\tablewidth{0pt}
\tablecaption{Fe~K$_{\alpha}$ Line and Collisional Ionization Emission Model
}
\label{tab:col}
\tablehead{
\colhead{Parameter} 
& \colhead{Value}  
& \colhead{Notes}  
}
\startdata
PL slope $\Gamma_1$ & $1.97^{+0.08}_{-0.06}$ & the slope below the break \\
PL norm [$10^{4}$] & $4.1^{+0.2}_{-0.2}$ & -- \\
\hline
Fe~K$_{\alpha}$ ${\rm N}_{\rm H}$ [$10^{24}~{\rm cm}^{-2}$] & $1.6$ & fixed \\
Fe~K$_{\alpha}$ norm. [$10^{4}$] & $2.7^{+0.1}_{-0.1}$ &  -- \\
Fe~K$_{\alpha}$ radius [$10^{4}~{\rm GM}/{\rm c}^{2}$] & $5.1^{+0.1}_{-0.3}$ & $4.1^{+11.5}_{-1.3}$ for $i=14^{+2}_{-2}$~deg \\
Fe~K$_{\alpha}$ inclination [degrees] & $13^{+74}_{-13}$ & convolution with ``Speith'' \\
\hline
CIE 1 kT [keV] & $7.0^{+1.2}_{-0.1}$ & frozen \\
CIE 1 ${\rm n}^{2}{\rm V}$ [$10^{62}~{\rm cm}^{-3}$] & $1.4^{+0.2}_{-0.2}$ & -- \\
CIE 1 $\sigma$ [${\rm km}~{\rm s}^{-1}$] & $810^{+310}_{-200}$ & -- \\
CIE 1 v [${\rm km}~{\rm s}^{-1}$] & $180^{+240}_{-60}$ & -- \\
\hline
CIE 2 kT [keV] & $20$ & fixed \\
CIE 2 ${\rm n}^{2}{\rm V}$ [$10^{62}~{\rm cm}^{-3}$] & $0.7^{+0.3}_{-0.3}$ & -- \\
CIE 2 $\sigma$ [${\rm km}~{\rm s}^{-1}$] & $230^{+230}_{-230}$ & -- \\
CIE 2 v [${\rm km}~{\rm s}^{-1}$] & $1500^{+200}_{-200}$ & -- \\
\hline
Flux ($10^{-11}~{\rm erg}~{\rm cm}^{-2}~{\rm s}^{-1}$)  & $2.6\pm 0.1$ & 0.0136--13.6~keV \\
Lum. ($10^{40}~{\rm erg}~{\rm s}^{-1}$) & $5.0\pm 0.2$  & 0.0136--13.6~keV \\
\hline
Cstat/$\nu$ & 2429.1/2308 & 1.7--10.7~keV \\
\enddata
\end{deluxetable}

\clearpage

\clearpage

\clearpage

\end{document}